\documentclass[aps,prb,preprint,groupedaddress,showpacs,amsmath,amssymb]{revtex4-2}

\usepackage{graphicx}
\usepackage{dcolumn}
\usepackage{bm}
\usepackage{hyperref}

\begin{document}

\title{Rapid suppression of charge density wave transition in LaSb$_2$ under pressure.}

\author{Sergey L. Bud'ko,$^{1,2}$ Shuyuan Huyan,$^{1,2}$ Paula Herrera-Siklody,$^{2}$ and Paul C. Canfield$^{1,2}$}

\affiliation{$^1$ Ames National  Laboratory, US DOE, Iowa State University, Ames, Iowa 50011, USA}
\affiliation{$^2$ Department of Physics and Astronomy, Iowa State University, Ames, Iowa 50011, USA}

\date{\today}

\begin{abstract}

LaSb$_2$ is found to be an example of an exceptionally pressure sensitive and tunable, two dimensional compound. In-plane electrical resistivity of LaSb$_2$ under pressure up to 12.9 kbar was measured in zero and applied magnetic field. The charge density wave transition (observed at $\sim 350$~K at ambient pressure) is completely suppressed by 6-7 kbar with significant (in comparison with the ambient pressure) increase in Fermi surface gapping and transition hysteresis just above ambient pressure.

\end{abstract}


\maketitle

\section{Introduction}
 LaSb$_2$ is a member of the light rare-earth diantimonides {\it R}Sb$_ 2$  ({\it R} = La-Nd, Sm) family in which the members all crystallize in the orthorhombic SmSb$_2$ structure. \cite{wan67a,hul79a} Single crystals of the members of the family were grown decades ago and basic physical properties were reported. \cite{can91a,bud98a} At that time \cite{bud98a} charge density wave (CDW) - like features in resistivity were observed (below 300 K) in PrSb$_2$ and NdSb$_2$, but not in other members of the family. LaSb$_2$ on the other hand is one of the first {\it stoichiometric} compounds where large, linear magnetoresistance was observed. {\cite{bud98a,you03a} Superconductivity with $T_c \approx 0.4 $~K was detected in LaSb$_2$, \cite{hul77a} and details of the superconducting state were studied at ambient and high pressure in Refs. \onlinecite{gal13a,guo11a} 

The possibility of CDW in LaSb$_2$ was discussed over the years. Quantum oscillations data and band structure calculations \cite{goo04a} were considered consistent with a conjecture of existence of CDW transitions, on the other hand, optical conductivity measurements \cite{dit11a} were interpreted as  ruling out the formation of a CDW phase above 20~K. Finally, signatures of CDW were discovered and studied in the La$_{1-x}$Ce$_x$Sb$_2$ substitutional series \cite{luc15a}, including a clear CDW-consistent feature in resistivity at $\sim 355$~K for pure LaSb$_2$. The observed suppression of CDW temperature with Ce - substitution (and the associated unit cell volume decrease) suggests the possibility of fair to moderate  pressure sensitivity  of this transition in pure LaSb$_2$.

This potential pressure sensitivity is experimentally addressed in this following by measurements of electrical resistivity of LaSb$_2$ in zero and applied magnetic field under pressure up to 12.9 kbar. We find that LaSb$_2$ is remarkably pressure sensitive with $T_{CDW}$ being suppressed to below 150 K by $\sim 4$~kbar and very likely fully suppressed by $\sim 6-7$~kbar. In addition we find that the resistive anomaly associated with $T_{CDW}$ changes dramatically when $T_{CDW}$ drops below the solidification temperature of the pressure media, suggesting that LaSb$_2$ is sensitive to even small deviations from purely hydrostatic pressure conditions. Finally we infer that for some pressure less than $\sim 2$~kbar there is a significant change in the degree of hysteresis in $T_{CDW}$ (changing from $\sim 2$ K at ambient pressure to more than 50 K above 2 kbar) and in the amount of Fermi surface that is gapped by the CDW transition (changing from roughly 10\%  at ambient pressure to $\sim 50$~\% above 2  kbar).

\section{Experimental details}
Large  single crystals of LaSb$_2$  were grown out of antimony flux.\cite{bud98a,can92a} Elemental La (99.99+ Ames Laboratory) and Sb (99.999+ Alfa Aesar) were combined in the ratio of La$_5$Sb$_{95}$ and placed into the bottom part of a 2 ml Canfield Crucible set (CCS). \cite{can16a,lsp} The CCS was sealed into an amorphous silica tube with silica wool on top of the CCS to act as a cushion during the decanting process. The sealed ampule was heated over 3 hours to 1100$^{\circ}$C, held at 1100$^{\circ}$C for 5 hours and then cooled to 1000$^{\circ}$C over 1 hour. After sitting at 1000$^{\circ}$C for 5 hours, the ampoule was cooled 675$^{\circ}$C over 99 hours. Upon reaching$^{\circ}$675 C, the ampule was removed from the furnace and decanted in a lab centrifuge. \cite{can20a} After cooling to room temperature, the ampule was opened, revealing large, sometimes crucible limited crystals of LaSb$_2$.  The crystals grow as soft plates with the $c$-axis perpendicular to the plates. The samples are metallic-micaceous, and the layers is malleable and readily deformed. Due to the high malleability of LaSb$_2$, a reasonable quality powder x-ray diffraction and even Laue diffraction are difficult to perform and were not attempted here. However, examination of the La - Sb binary phase diagram \cite{mas92a} as well as x-ray diffraction data \cite{luc15a} on crystals grown in a similar way show unambiguously that the resulting crystals are indeed orthorhombic LaSb$_2$. 

Standard, linear four-probe ac resistance at ambient pressure and under pressure was measured on bar - shaped samples in a $I || ab$ geometry. The 0.25 $\mu m$ diameter Pt wires were spot-welded to the sample and then the contacts were covered with Epo-Tek H20E silver-filled epoxy for better mechanical stability. The contact resistance values were $\sim 1 \Omega$ or lower. The frequency used was 17 Hz, typical current values were 3-5 mA. Magnetoresistance was measured in a transverse configuration, $H || c$, $I || ab$. The measurements were performed using the ACT option of a  Quantum Design Physical Property Measurement System (PPMS) instrument. Pressure was generated  in a hybrid, BeCu / NiCrAl piston - cylinder pressure cell (modified version of the one used in Ref. \onlinecite{bud86a}). A 40 : 60 mixture of light mineral oil and n-pentane was used as a pressure-transmitting medium. This medium solidifies at room temperature in the pressure range of 30 - 40 kbar, \cite{bud86a,kim11a,tor15a} which is above the maximum pressure in this work. Elemental Pb was used as a low temperature pressure gauge.\cite{eil81a}  The measurements were performed both on increase and decrease of pressure. Given that the pressure inside of a piston cylinder cell changes with temperature, we used the results of Ref. \onlinecite{xia20a} to (approximately) incorporate this difference in the cited pressure values. In the following we will use $P_{LT}$ for the low temperature pressure values obtained using Pb and  $P$ for recalculated  pressure values at higher temperatures that are used in the $P - T$ and related plots shown in Figs. \ref{F4} and \ref{F5} below. It should be noted we only present finite pressure data that has $P_{LT} > 0$}.

\section{Results}

Resistivity measurements up to 380 K at ambient pressure are shown in Fig. \ref{F1}. In agreement with Ref. \onlinecite{luc15a}, a clear feature in resistivity resembling CDW transition is observed at $\sim 350$~K. This transition has small but resolvable hysteresis of $\sim 2$~K. 

An example of the resistivity data taken under pressure ($P_{LT} = 0.1$~kbar is presented in Fig. \ref{F2}. Several features are noteworthy: (i) CDW transition is shifted to lower temperatures in comparison to ambient pressure; (ii) the hysteresis is increased significantly, from $\sim 2$~K at ambient pessure to about 50 K; (iii) the feature in resistivity associated with the CDW transition became significantly more pronounced,  changing from a $\sim 10$~\% increase in resistance at ambient pressure to a nearly 100\% increase at 0.1 kbar; (iv) there is clear, low temperature magnetoresistance with a local minimum in $\rho(T)$ appearing near 40 K (such local minimum has been observed in resistivity of a number of semi-metals in magnetic field \cite{mun12a,jon17a} and is understood within the Kohler's rule physics \cite{pip89a});  and (v) 140 kOe magnetic field does not affect the CDW transition temperature (Fig. \ref{F2}(b), inset). 

This last observation allows us to shorten the measurements time by evaluating the CDW transition hysteresis under pressure from comparison of $H = 0$ data on cooling and $H = 140$~kOe on warming. A subset of such data is shown in Fig. \ref{F3}. The resistivity at 300 K decreases slowly and monotonically under pressure. For $0.1~\text{kbar} \leq P_{LT} \leq 3.6$~kbar the feature associated with the CDW transition shifts to lower temperatures and gradually decreases in its amplitude. The $T_{CDW}$ continues to decrease for $4.5~\text{kbar} \leq P_{LT} \leq 5.5$~kbar, but the feature broadens and becomes barely resolved. Finally, no features in resistivity are observed in the $8.6~\text{kbar} \leq P_{LT} \leq 12.9$~kbar pressure range, so by or before 8.6 kbar the CDW transition is completely suppressed. The suppression of CDW transition is reflected in the base temperature magnetoresistance as well, even if in a subtle way. 

\section{Discussion and Summary}

\begin{figure}
\begin{center}
\includegraphics[angle=0,width=120mm]{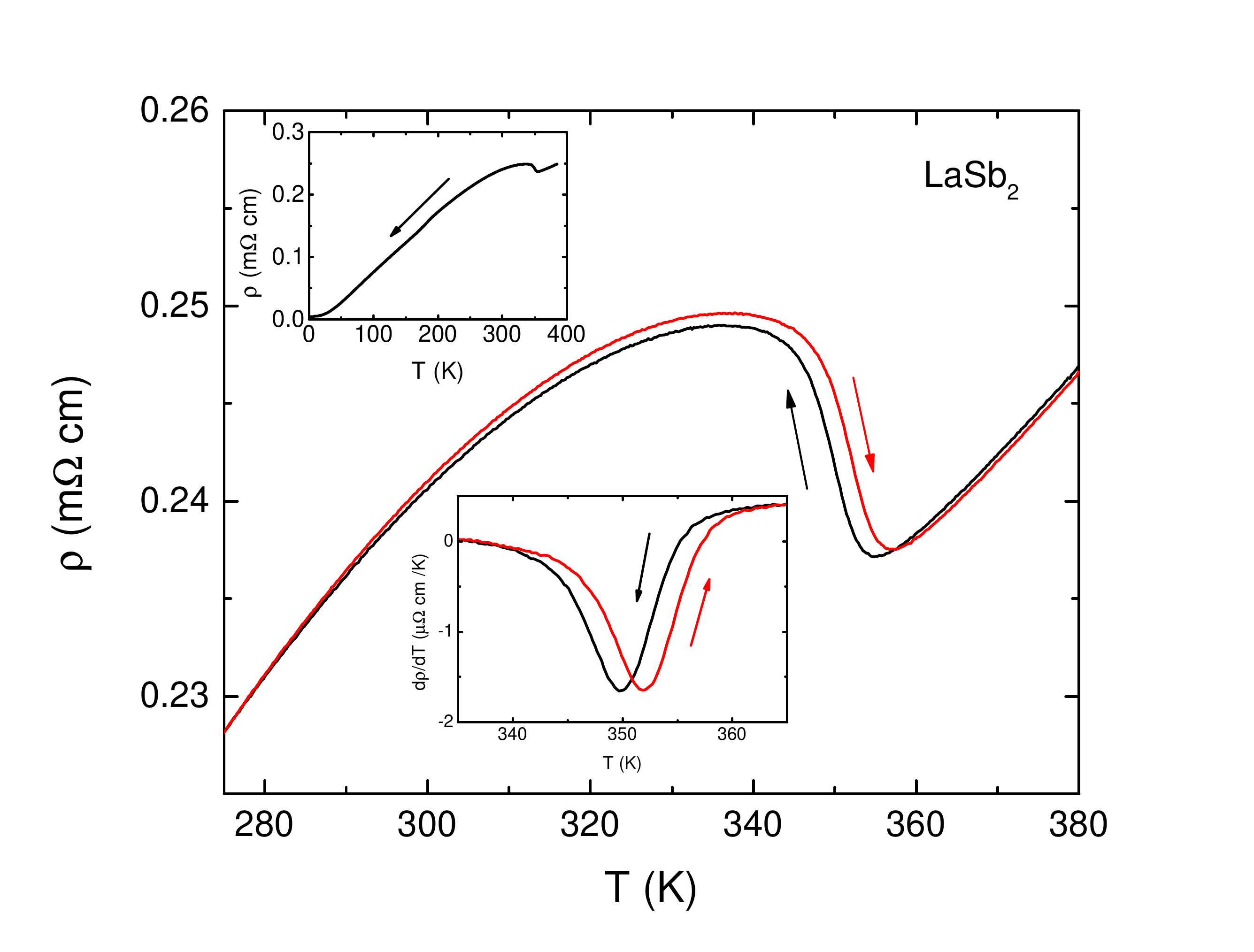}
\end{center}
\caption{(color online) In-plane resistivity measurements of LaSb$_2$ on warming and cooling between 275 K and 380 K. Upper inset - measurements on cooling in he whole temperature range. Lower inset - derivatives, $d \rho /dT$, in the region of the proposed CDW transition. Note, the sample for these measurements is taken from the same batch as the sample measured under pressure. } \label{F1}
\end{figure}

\begin{figure}
\begin{center}
\includegraphics[angle=0,width=120mm]{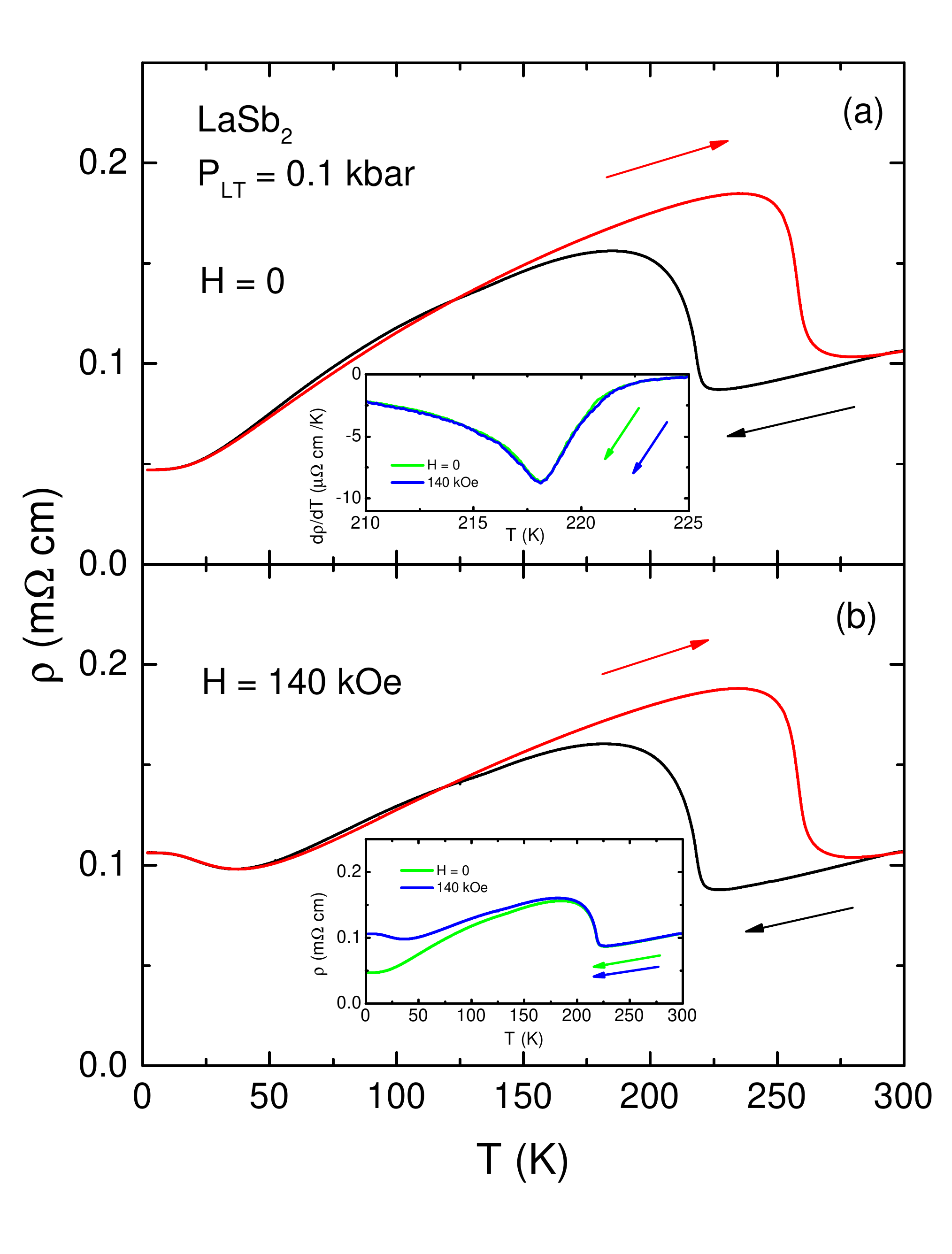}
\end{center}
\caption{(color online) Resistivity of LaSb$_2$ at $P_{LT} = 0.1$~kbar on warming and cooling: (a) in zero applied field, (b) for $H = 140$~kOe. Inset to (a): derivatives $d \rho / dT$ at temperatures close to the transition for the data taken on cooling in $H = 0$ and $H = 140$~kOe.  Inset to (b): $\rho(T)$ data taken on cooling in $H = 0$ and $H = 140$~kOe. } \label{F2}
\end{figure}

\begin{figure}
\begin{center}
\includegraphics[angle=0,width=120mm]{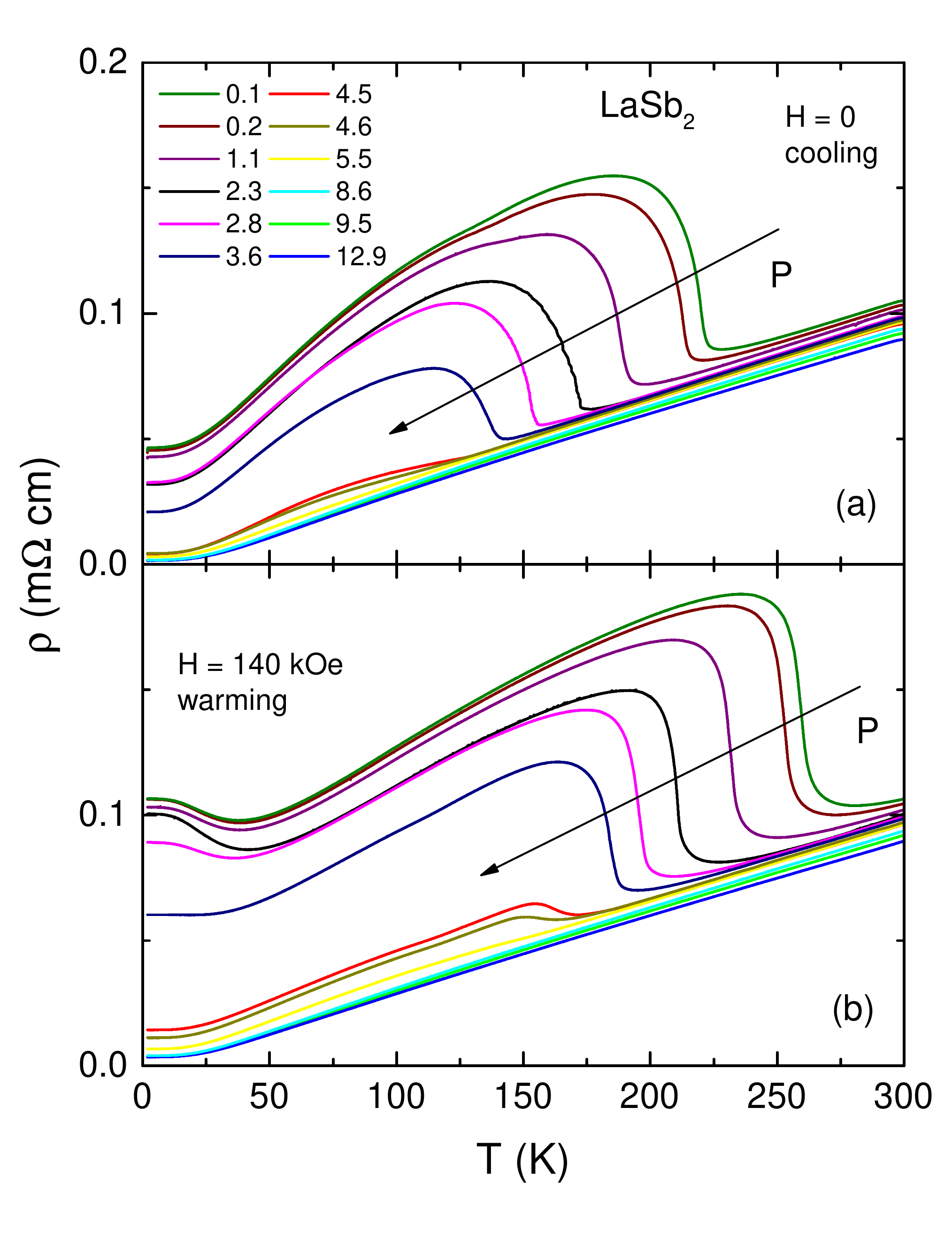}
\end{center}
\caption{(color online) Subset of the resistivity data of LaSb$_2$ under pressure: (a) $H = 0$, cooling, (b) $H = 140$~kOe, warming. Arrows point the direction of pressure increase. The numbers in the legend are low temperature pressure value, $P_{LT}$, in kbar. } \label{F3}
\end{figure}

\begin{figure}
\begin{center}
\includegraphics[angle=0,width=120mm]{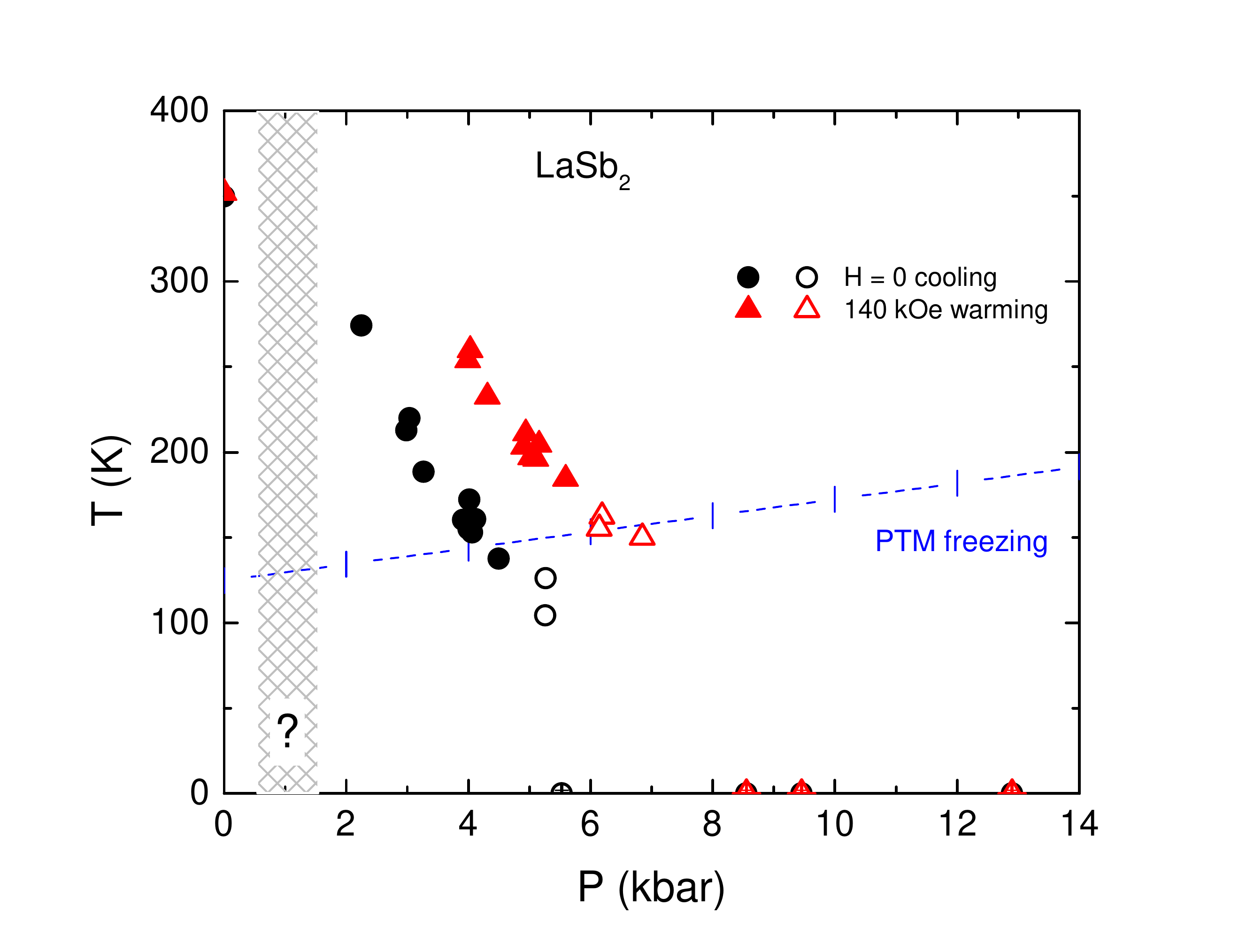}
\end{center}
\caption{(color online) $P - T$ phase diagram of LaSb$_2$ inferred from the resistivity data. Symbols: CDW transitions (using minimum in $d \rho / dT$ criterion), circles - on cooling,  triangles - on warming. See the text for the difference between filled and open symbols.  Crosshatched area indicates hypothetical electronic / structural transition.  Dashed line corresponds to 40 : 60 mineral oil - n-pentane  pressure transmitting medium (PTM) freezing line. \cite{tor15a}. See the text for more details.  } \label{F5}
\end{figure}

The temperature-dependent resistivity data taken at different pressures allows for the construction of a $P - T$ phase diagram for LaSb$_2$ presented in Fig. \ref{F5}. The symbols correspond to CDW transition temperatures at different pressures as determined from the minima in the derivatives $d \rho / dT$ of the electrical transport data taken on cooling and on warming, as illustrated in the insets of Figs. \ref{F1} and \ref{F2}. The open and closed symbols on the T-P phase diagram are associated with the abrupt decrease (and broadening) of the CDW-like feature in resistivity for the $4.5~\text{kbar} \leq P_{LT} \leq 5.5$~kbar datasets (Fig. \ref{F3}). Although it would be tempting to conject the presence of a transition in the 3-4 kbar pressure range, a more mundane explanation seems to be plausible. The CDW-like transition has unambiguous hysteresis pointing to its first order. It is likely that it is accompanied by (anisotropic) changes in the lattice parameters and volume. If this is hypothesis is correct, then we might expect that the experimental  signature of the transition crucially depends on the level of hydrostaticity of the pressure transmitting medium (PTM) (for example, above and below of the medium freezing line, CaFe$_2$As$_2$ under pressure \cite{tor08a,yuw09a,can09a} is one of the recognized examples of such sensitivity). Indeed, the position of the freezing line of the PTM used in our measurements \cite{tor15a} (Fig. \ref{F5}) is in agreement with the pressure-temperature range where the abrupt change of the CDW-like signature in resistivity is observed. Clearly, measurements in He gas / liquid pressure medium would be required to address the issue of hydrostaticity.

\begin{figure}
\begin{center}
\includegraphics[angle=0,width=120mm]{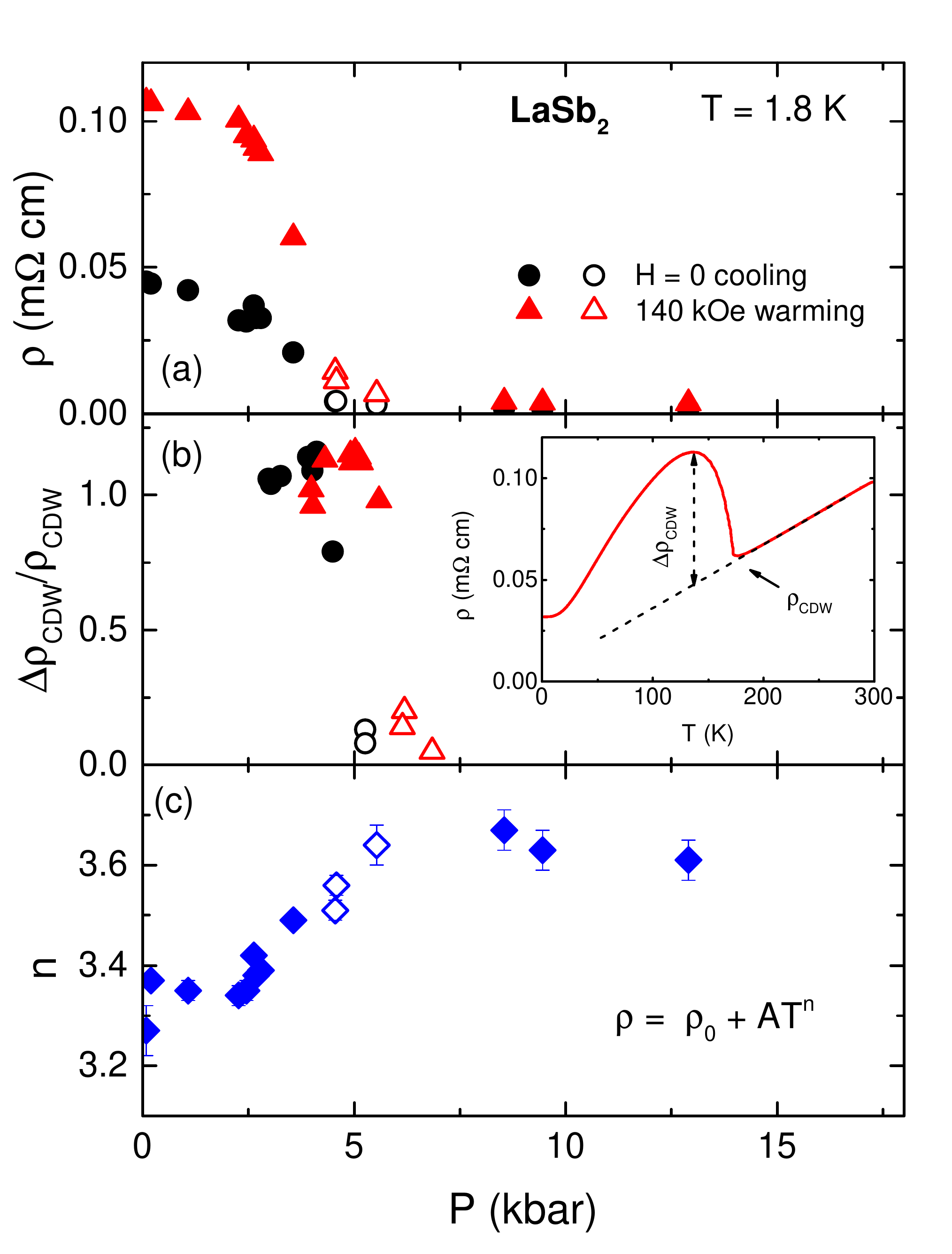}
\end{center}
\caption{(color online) (a) Resistivity at $T = 1.8$~K in zero and 140~kOe applied magnetic field as a function od pressure. (b) Relative size of the feature in resistivity associated with CDW, $\Delta \rho_{CDW} / \rho_{CDW}$, as a function of pressure. the inset shows how  $\Delta \rho_{CDW}$ and  $\rho_{CDW}$ were defined. (c) Exponent $n$ in low temperare fit of $H = 0$ resistivity $\rho = \rho_ + AT^n$. See the text for the difference between filled and open symbols.} \label{F4}
\end{figure}

To further explore the effects of applied pressure on LaSb$_2$, in figure \ref{F4} we present the pressure dependence of the 1.8 K resistivity and the relative change in the resistivity associated with CDW transition , $\Delta \rho_{CDW} / \rho_{CDW}$,  which can serve as a simple caliper of how much of the Fermi surface is gapped as a result of the CDW transition. \cite{tor07a}. In addition, we plot the pressure dependence of the exponent $n$ in the fit of the zero field, low temperature ($ 1.8~\text{K} \leq T \leq 15~\text{K}$ ) resistivity to a power law, $\rho = \rho_0 + AT^n$,  as well in Fig. \ref{F4}(c). For each of the plots we adopt the convention we used in figure \ref{F5} and show the data associated with finite $T_{CDW} < T_{solidification}$ as open symbols. Note that if $T_{CDW} = 0$~ K, then there is no signature of a CDW transition and there is no broadening of (nonexistent) features.

The base temperature resistivity (Fig. \ref{F4}(a)) decreases with increasing rapidity for pressure up to $\sim 5-6$~kbar, then the behavior changes abruptly and becomes only slightly pressure-dependent at higher pressures. This behavior is not unexpected, since below the pressure at which CDW is driven to $T = 0$~K an additional contribution from suppression of the resistive increase associated with the Fermi surface gapping due to CDW plays an important role. Similar behavior was observed e.g. in LaAu$_{0.970}$Sb$_2$ under pressure. \cite{xia20b}

The behavior of $\Delta \rho_{CDW} / \rho_{CDW}$ parameter is more complex. At ambient pressure it is rather small, suggesting, within a very simple model, that $\lesssim 10 \%$ of the Fermi surface  is gapped. On pressure increase to $\sim 2$~kbar (at $T_{CDW}$) $\Delta \rho_{CDW} / \rho_{CDW}$  jumps to $\sim 1$, that implies gapping of roughly half of the Fermi surface. Then, in 5-7 kbar range, $\Delta \rho_{CDW} / \rho_{CDW}$ abruptly decreases, whereas at 8.6 kbar and above this feature is not observed any more.

The exponent $n$ ranges from $\sim 3.3$ to $\sim 3.7$ as pressure is increased, initially varying only weakly as $T_{CDW}$ is weakly suppressed. As $T_{CDW}$ is suppressed more rapidly, $n$  increases from 3.3 to 3.7. Once $T_{CDW}$ is suppressed to zero, $n(P)$ flattens.

It is noteworthy that whereas the low temperature resistivity ($\rho_{1.8~\text{K}}$, $n$) appears to be not very sensitive to $T_{CDW}(P)$ crossing $T_{solidification}(P)$ line, the changes in $\Delta \rho_{CDW} / \rho_{CDW}$ on this crossing are abrupt and unambiguous. This is actually not too surprising if the effects of non-hydrostaticity are thought to be a broadening of the resistive feature of the transition rather than a significant change in the low temperature state. If $T_{CDW}$ is well above base temperature or the $T \leq 15$~K range of power law fitting, then the affects of this broadening will be long past, at higher temperatures. On the other hand, the$\Delta \rho_{CDW} / \rho_{CDW}$ is inherently associated with the $T_{CDW}$ temperature region.

As mentioned above, the ambient pressure data (Fig. \ref{F1}) and the data for $0.1~\text{kbar} \leq P_{LT} \leq 3.6$~kbar (Fig.\ref{F3}) show very clear, CDW - like feature in resistivity, however comparing those (i) the hysteresis between data taken on cooling and on warming is notably different; (ii) the size of the CDW - like anomaly is significantly different. We believe that neither some difference in the in-plane direction of the current, nor possible temperature lag due to the thermal mass of the cell (rates of $\sim 0.3$~K/min were used) could explain these drastic differences. (Note, that at ambient pressure similar relative size of the anomaly was observed in Ref. \onlinecite{luc15a}, and no abrupt change of hysteresis under pressure in the same or similar cell and measurements protocol was reported in Ref. \onlinecite{bud20a}). This difference between ambient pressure and small applied pressure is fully consistent with the sensitivity of the CDW feature to hydrostatic / non-hydrostatic conditions discussed above. We speculate that (i)  LaSb$_2$ mignt have very low pressure, significantly non-linear compressibility (not too surprising for van der Waals - like structure) that leads to rapid change of nesting, or (ii) there might be a dramatic pressure-induced change of electronic structure near the Fermi level  (possible caused by some subtle structural transformation) at low pressure that causes these striking difference in the degree of nesting and thermal hysteresis. This possible transition is marked by the crosshatched area in the $P - T$ phase diagram (Fig. \ref{F5}). The complex (calculated)  Fermi surface of LaSb$_2$ \cite{goo04a} is not inconsistent with the possibility of significant change of nesting under moderate pressure.

Regardless of possible complications caused by PTM freezing, and possible change of electronic structure  in 1 kbar range, the data show that the CDW transition in LaSb$_2$  is very fragile, so that 6-7 kbar pressure is enough to suppress $T_{CDW}$ to $T = 0$~K, this corresponds to overall pressure derivative of $\sim - 50$~K/kbar, that is more than an order of magnitude larger in the absolute value than that for LaAgSb$_2$ \cite{tor07a} ($\approx - 4.3$~K/kbar). This might be due some specific details of the band structure of these materials, as well as to possible differences in elastic modulae.

\begin{figure}
\begin{center}
\includegraphics[angle=0,width=120mm]{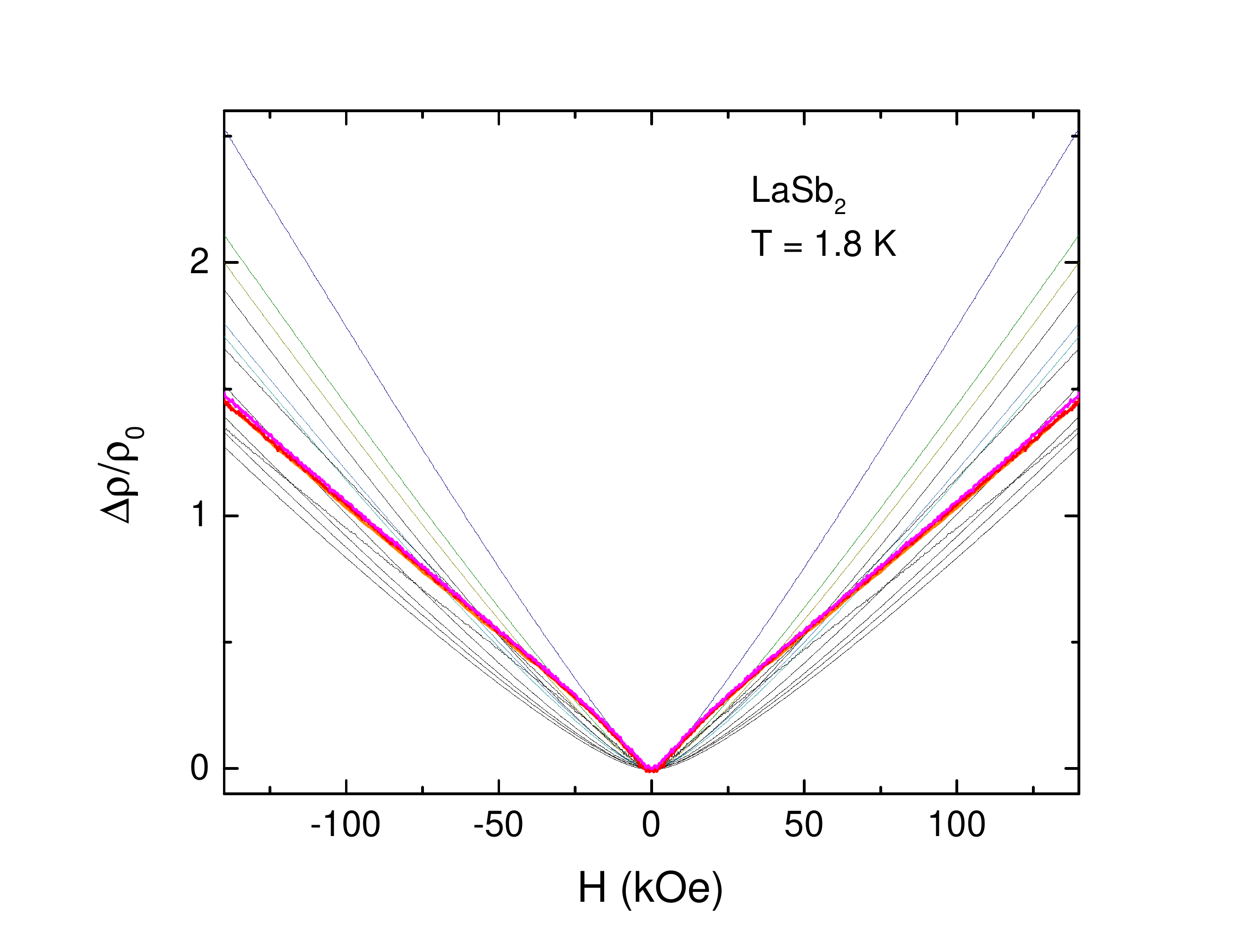}
\end{center}
\caption{(color online) Base temperature magnetoresistance of LaSb$_2$ at different pressures. Thin lines: data for   $0.1~\text{kbar} \leq P_{LT} \leq 5.5$~kbar. Thicker (magenta. orange, and red)  lines - data for 8.6, 9.5 and 12.9 kbar. Since there is a small Hall contribution to the data, this plot show even in magnetic field resistivity component using $\rho (H) = 1/2 (\rho(+H) + \rho(-H))$. } \label{F6}
\end{figure}

A clear change in the base temperature magnetoresistance (MR) (figure \ref{F6}) can be seen when pressures exceed the 6-7 kbar range. For $P_{LT}$ = 8.6, 9.5 and 12.9 kbar a very clear, low field negative curvature / or shoulder develops around $H \sim 25$~kOe. For lower pressures the MR has a positive curvature over the whole applied field range. These data are consistent with a change in the band structure associated with the loss of the lower pressure CDW state. Given that MR is traditionally plotted as $\Delta \rho/\rho_0$ and given that $\rho_0$ changes rapidly and by very large amounts (due the changing nature of the CDW gapping) the lower pressure variation of the MR is non-monotonic. This said, the functional change in the magnetic field - dependent MR for $P_{LT} > 7$~ kbar is clear and unambiguous. 

Our data on CDW suppression in LaSb$_2$ could be compared with the evolution of superconductivity in LaSb$_2$ under pressure in this material. \cite{guo11a} The $T_c$ reportedly increases under pressure, the superconducting dome of "3D superconductivity"  has a maximum at about 4-5 kbar and then $T_c$ decreases with further pressure increase, but at a measurably slower rate. Qualitatively, such behavior appear to be consistent with a simple picture of coexistence and competition between CDW and superconductivity. \cite{bil76a,bal79a}

To summarize, the CDW state in LaSb$_2$ was found to be quickly (by 6-7 kbar) supressed under pressure with the overall pressure derivative of $\sim - 50$~K/kbar . We hypothesize that (i) the CDW-like transition has a structural component that causes enhanced sensitivity of CDW transition to the level of pressure hydrostaticity, and (ii) that there is either a very low pressure, non-linear compressibility that leads to rapid change of nesting or band structure very close to pressure induced change that dramatically changes degree of nesting of the CDW. All of these data and observations point out that LaSb$_2$ is a long overlooked, highly pressure sensitive, low dimensional material.

\begin{acknowledgments}

We would like to thank Hermann Suderow for useful communications. Work at the Ames National Laboratory was supported by the U.S. Department of Energy, Office of Science, Basic Energy Sciences, Materials Sciences and Engineering Division. The Ames National Laboratory is operated for the U.S. Department of Energy by Iowa State University under contract No. DE-AC02-07CH11358. PCC and SLB wish to acknowledge Hidalgo Alonso Quijano for inspiring this line of inquiry.

\end{acknowledgments}

\clearpage

\clearpage

\clearpage

\clearpage

\clearpage

\clearpage

\end{document}